\documentclass[onecolumn,showpacs,aps,floatfix,pra,superscriptaddress,nofootinbib]{revtex4}
\usepackage{color}
\usepackage{graphicx}
\usepackage{bm}
\usepackage{amsmath}
\begin{document}
\title{Effect of quantum statistics on the gravitational weak equivalence principle}
\author{S. V. Mousavi}
\email{vmousavi@qom.ac.ir}
\affiliation{Department of Physics, The University of Qom, Qom, Iran}
\affiliation{School of Physics, Institute for Research in Fundamental Sciences (IPM), P.O.Box 19395-5531, Tehran, Iran }

\author{A. S. Majumdar}
\email{archan@bose.res.in}
\affiliation{S. N. Bose National Center for Basic Sciences, Block JD, Sector III, Salt Lake, Kolkata 700 098, India}

\author{D. Home}
\email{dhome@bosemain.boseinst.ac.in}
\affiliation{ CAPSS, Department of Physics, Bose Institute, Sector-V, Salt Lake, Kolkata 700 091, India}

%%%%%%%%%%%%%%%%%%%%%%%%%%%%%%%%%%%%%%%%%%%%%%%%%%%%%%%%%%%%%%%%%%%

%
\begin{abstract}

We study the effect of quantum statistics on the arrival time
distribution of quantum particles computed through the probability
current density. It is shown that symmetrization or asymmetrization
of the wave function affects the arrival time distribution for even
freely propagating particles. In order to investigate the effect of
statistics on the weak equivalence principle in quantum mechanics
(WEQ), we then compute the mean arrival time for wave packets in
free fall. The violation of WEQ through the effect of statistics on
the mass-dependence of the mean arrival time is clearly exhibited.
We finally evaluate the effect of spin on the violation of WEQ using
a different approach by including an explicit spin-dependence in the
probability current distribution, and compare it with the approach
using particle statistics. Our results show WEQ re-emerges smoothly
in the limit of large mass.

\end{abstract}

 \maketitle

%%%%%%%%%%%%%%%%%%%%%%%%%%%%%%%%%%%%%

\section{Introduction}

The statistics of elementary particles is a key underlying
ingredient of quantum field theory which has been most successful in
explaining three, {\it viz.} electromagnetic, weak and strong
nuclear forces, of the four fundamental forces in nature. On the
other hand, the explicit role of quantum statistics in physical
phenomena such as Bose Condensation \cite{be} is also being
increasingly displayed in recent years. Practical applications of
such statistics is potentially wide, ranging from optical lattices
\cite{ol} to quantum information \cite{inf}. In most practical
manifestations of the role of Bose statistics, a crucial role is
played by the measurement of the time of flight distribution
\cite{tof1}. For example, the coherence properties of a condensate
of ultracold atoms is imprinted on the time of flight distribution
when the trapping potential is released and the particles undergo
free fall. Thus, the effect of gravity on the motion of quantum
particles seems to play an important role in the inference of
attributes related to the statistics obeyed by them.

The motion of freely falling particles is traditionally taken to
conform to the weak equivalence principle (WEP) of gravitation which
states that all sufficiently small objects fall with the same
acceleration independent of their mass and constituent in a uniform
gravitational field. WEP is regarded to be a fundamentally classical
and local principle. Study of the equivalence principle in quantum
mechanics has evoked enough interest. A statement of the principle
in quantum mechanics is as follows: "The results of experiments in
an external potential comprising just a sufficiently weak,
homogeneous gravitational field, as determined by the wavefunction,
are independent of the mass of the system" \cite{Ho_book_1993}. This
assertion is also called the weak equivalence principle of quantum
mechanics (WEQ).

Different approaches have been used to study the possibility of
violation of weak equivalence principle in quantum mechanics, starting
from the prediction of mass-dependence of the
radii, frequencies and binding energy of a particle in an
external gravitational field \cite{Greenberger}.
Experimental probes into possibility of violation of the weak
equivalence principle in quantum mechanics have been attempted using
interference phenomenon associated with the gravitational potential
in neutron and atomic interferometry experiments
\cite{CoOvWe-PRL-1975, PeChCh-nature-1999}.
However, the phase shifts
observed in such experiments \cite{CoOvWe-PRL-1975} may be expressed in terms of directly
observable quantities which turn out to be independent of mass \cite{La-GRG-1996},
thus ruling out the violation  in this context. Moreover, atomic
interferometry experiments \cite{PeChCh-nature-1999} need to be performed with different atomic
species in order for a rigorous experimental test of the weak equivalence
principle at the quantum level.
A gedanken experiment studying the free fall of quantum
test particles in a uniform gravitational field predicts
mass-dependence of the time of flight distribution
\cite{ViOn-PRD-1997}. Another approach using a model quantum clock
predicts violation of WEQ in the vicinity of the turning point of
classical trajectories \cite{Da-CQG-2004}.
An explicit mass dependence of the position probabilities has been shown for
quantum particles projected upwards against gravity around both the
classical turning point and the point of initial projection using
Gaussian \cite{AlMaHoPa-CQG-2006} and non-Gaussian
\cite{ChHoMaMoMoSi-CQG-2012} wavepackets. Moreover, explicit mass
dependence of the mean arrival time at an arbitrary detector
location has also been predicted for a Gaussian
\cite{AlMaHoPa-CQG-2006} wave-packet under free fall. Such
mass-dependence may be enhanced using suitably chosen non-Gaussian
wavepackets \cite{ChHoMaMoMoSi-CQG-2012}. Thus, both the position
probabilities and the mean arrival time show the violation of WEQ
for quantum particles described by Gaussian and non-Gaussian
wavepackets. It has been discussed violation of WEQ arises as a
consequence of the spread of wave packets, the magnitude of which
itself depends on the mass. An illustration of this effect has been
presented based on Bohmian trajectories
\cite{ChHoMaMoMoSi-CQG-2012}.

The above studies have established the violation of WEQ in single
particle quantum mechanics. However, the effect of quantum
statistics on the violation of WEQ has remained  uninvestigated till
date. As pointed out earlier, several important phenomena based on
quantum statistics is experimentally revealed through the
measurement of time of flight of quantum particles in free fall.
Consideration of quantum mechanical effects on such time of flight
distributions beyond the standard semi-classical analysis could
reveal interesting observational effects, as has been debated
earlier in the literature \cite{tof2}. In the present work we are
interested to examine the effect of quantum statistics on the WEQ.
With the above motivation in this paper we  study the effect of
statistics on the arrival time distribution of a system of freely
falling wavepackets consisting of two identical particles.

Our analysis is based on the probability current approach for
computing the mean arrival time distribution of wavepackets
\cite{leavens}. Before embarking on the study of WEQ, we will first
consider the free evolution of a system of two identical particles
to investigate if there is any effect of Bose-Einstein (BE) or
Fermi-Dirac (FD) statistics on the arrival time distributions.
Quantum statistics is fundamentally related to the spin of the
quantum particles. It has been shown earlier that inclusion of spin
enables a unique determination of the relativistic probability
current \cite{Ho-PRA-1999,fink} leading to an explicit
spin-dependent effect at the non-relativistic level for freely
evolving fermions \cite{spin1} as well as bosons \cite{struyve}.
Here we present a calculation of the effect of such explicit
spin-dependence on the violation of WEQ for a system of two
identical particles in free fall.

The plan of this paper is as follows. In the next section we present
the framework for studying the arrival time distribution for a
system of two particles using the probability current approach. We
show that symmetrization or anti-symmetrization of the wavefunction
leads to differences from the case of Maxwell-Boltzmann statistics
when there is overlap of the wavepackets.
 In section III the effect of BE and FD statistics on freely propagating
wavepackets is studied. In section IV we address the main motivation
of the paper by studying the violation of WEQ separately using two
different schemes, {\it viz}, (a) effect of statistics on the
arrival time distribution, and (b) effect of explicit
spin-dependence on the arrival time distribution of freely falling
wavepackets. Section V contains a summary of the main results along
with certain concluding remarks.
%

%===================================================

\section{Arrival time distribution for a two-particle system}

Consider a two-body system composed of two non-interacting identical
particles in an external field. Identical particles are classically
distinguishable and obey Maxwell-Boltzmann (MB) statistics, while
they are indistinguishable in quantum mechanics and obey different
statistics. For Fermi-Dirac (Bose-Einstein) statistics the total
wavefunction must be antisymmetrized (symmetrized) under the
exchange of particles in the system. Since particles do not
interact, solutions of the Schr\"{o}dinger equation are constructed
from two single-particle wavefunctions $\psi_a$ and $\psi_b$ as
follows \cite{Ho_book_1993}:
\begin{eqnarray}
\Psi_{\mathrm{MB}}(z_1, z_2, t) &=& \psi_a(z_1, t) \psi_b(z_2, t)~,\\
\Psi_{\pm}(z_1, z_2, t) &=& N_{\pm}[\psi_a(z_1, t) \psi_b(z_2, t)
\pm \psi_b(z_1, t) \psi_a(z_2, t)]~,\label{eq: two-body psi}
\end{eqnarray}
where upper (lower) sign stands for BE (FD) statistics and the normalization constants are
given by $N_{\pm} = [2 ( 1 \pm |\langle \psi_a(t) | \psi_b(t) \rangle|^2 )]^{-1/2} $.

Two-body wavefunctions (\ref{eq: two-body psi}) are solutions of the
two-body Schr\"{o}dinger equation
\begin{eqnarray} \label{eq: two-body Sch}
 i\hbar \frac{\partial}{\partial t} \Psi(z_1, z_2, t) &=& \left(
-\frac{\hbar^2}{2m} \frac{\partial^2}{\partial z_1^2}
-\frac{\hbar^2}{2m} \frac{\partial^2}{\partial z_2^2} + V(z_1, t) +
V(z_2, t) \right) \Psi(z_1, z_2, t)~,
\end{eqnarray}
where the Hamiltonian is clearly symmetric under the exchange of
$z_1$ and $z_2$.

The one-body probability density for observing a particle at point
$z$ regardless of the position of the other particle, is given by
\cite{Zelev-book-2011}
\begin{eqnarray} \label{eq: rho1}
\rho_1(z, t) &=& \int dz_1 dz_2~\delta(z-z_1)~|\Psi(z_1, z_2, t)|^2
= \int dz_2~|\Psi(z, z_2, t)|^2
\nonumber \\
&=&
|N_{\pm}|^2 \int dz_2 ~ \bigg| \psi_a(z, t) \psi_b(z_2, t) \pm \psi_b(z, t) \psi_a(z_2, t) \bigg|^2
\nonumber \\
& \equiv &
|N_{\pm}|^2 \bigg( |\psi_a(z, t)|^2 \pm
2~\Re\left[ \langle \psi_a(t) | \psi_b(t) \rangle ~ \psi^*_b(z, t)
\psi_a(z, t) \right]  + |\psi_b(z, t)|^2 \bigg)~,
\end{eqnarray}
where $\langle \psi_a(t) | \psi_b(t) \rangle = \int dx ~ \psi^*_a(x,
t)  \psi_b(x, t) $ is the overlap integral and the normalization
condition for single-particle wavefunctions $\psi_a$ and $\psi_b$
has been used. From the unitarity property of the time evolution
operator, $ U(t,0) = e^{-i H t/\hbar}$, one sees that the overlap
integral is independent of time, i.e., $ \langle \psi_a(t) |
\psi_b(t) \rangle = \langle \psi_a(0) | U^{\dagger}(t, 0) U(t,0) |
\psi_b(0) \rangle = \langle \psi_a(0) | \psi_b(0) \rangle$.

The single-particle continuity equation may be derived
\cite{MoMi-PS-2014} in the following way. The partial derivative of
$\rho_1(z, t)$ with respect to the time is calculated from eq.
(\ref{eq: rho1}) by performing integration by parts and using the
fact that the wavefunction vanishes at infinity, to obtain
\begin{eqnarray}
\frac{\partial \rho_1(z, t)}{\partial t} &=& -\frac{\hbar}{m} \Im
\left \{ \frac{\partial}{\partial z} \int dz_2~ \Psi^*(z, z_2, t)
\frac{\partial \Psi(z, z_2, t)}{\partial z} \right \} ~,
\end{eqnarray}
Now, using the normalization condition for the single-particle wavefunctions,
one obtains the  continuity equation \cite{MoMi-PS-2014}
\begin{eqnarray} \label{eq: con_eq}
\frac{\partial \rho_1(z, t)}{\partial t} + \frac{\partial}{\partial
z} j_1(z, t) &=& 0~,
\end{eqnarray}
for the one-body probability density $\rho_1(z, t)$ (\ref{eq: rho1}),
with the  one-body probability current density given by
\begin{eqnarray} \label{eq: prob_qm_twobody}
j_1(z, t) &=& \frac{\hbar}{m} |N_{\pm}|^2 \Im \left \{ \psi^*_a
\frac{\partial \psi_a}{\partial z} + \psi^*_b \frac{\partial
\psi_b}{\partial z} \pm \langle \psi_a(t) | \psi_b(t) \rangle ~
\psi^*_b \frac{\partial \psi_a}{\partial z} \pm \langle \psi_b(t) |
\psi_a(t) \rangle ~ \psi^*_a \frac{\partial \psi_b}{\partial z}
\right \}~.
\end{eqnarray}

For distinguishable particles obeying classical MB statistics, each
particle has its corresponding continuity equation,
\begin{eqnarray}
\frac{\partial \rho^{(i)}(z, t)}{\partial t} +
\frac{\partial}{\partial z} j^{(i)}(z, t) &=& 0~,
\end{eqnarray}
with
\begin{eqnarray}
\rho^{(1)}(z, t) = |\psi_a(z, t)|^2~,
~~~~~~~~~~~~~~~~~~\rho^{(2)}(z, t) = |\psi_b(z,
t)|^2~,
\end{eqnarray}
and
\begin{eqnarray}
j^{(1)}(z, t) = \frac{\hbar}{m} \Im \left \{
\psi^*_a(z, t) \frac{\partial \psi_a(z, t)}{\partial z} \right \} ~,
~~~~~~~~~~~ j^{(2)}(z, t) = \frac{\hbar}{m} \Im \left
\{ \psi^*_b(z, t) \frac{\partial \psi_b(z, t)}{\partial z} \right
\}~.
\end{eqnarray}
So, for a two-particle system, the MB probability density and probability
current density are given by
\begin{eqnarray}
\rho_{\text{MB}}(z, t) = \frac{1}{2} \left(\rho^{(1)}(z, t) +
\rho^{(2)}(z, t) \right) ~, ~~~~~~~~~~~j_{\text{MB}}(z, t) =
\frac{1}{2} \left(j^{(1)}(z, t) + j^{(2)}(z, t) \right)
\end{eqnarray}

From eqs. (\ref{eq: rho1}) and (\ref{eq: prob_qm_twobody}), one sees
that as long as the single-particle wavefunctions have negligible
overlap, i.e., $ \langle \psi_a(t) | \psi_b(t) \rangle \simeq 0$,
there is no need for symmetrization. As a result, one can ignore
indistinguishability of particles and thus use MB statistics for
which motions of particles are independent. Hence, in the latter
case a similar analysis to the one-body systems discussed earlier
\cite{AlMaHoPa-CQG-2006, ChHoMaMoMoSi-CQG-2012} may be employed.

The probability current approach for computation of the arrival time
distribution of quantum particles not only provides an unambiguous
definition of arrival time at the quantum mechanical level
\cite{leavens,Ho-PRA-1999,fink, spin1}, but also furnishes a way of
obtaining the classical limit of the mean arrival time  for massive
quantum particles \cite{classlim}. Here we employ the probability
current approach to study the effect of particle statistics on the
arrival time distribution of a two-body system. In this approach,
the arrival time distribution at a detector location $z = Z$ is
given by \cite{leavens},
\begin{eqnarray} \label{eq: arrival distribution}
\Pi(Z, t) &=& \frac{ |j_1(Z, t)| }{\int_0^{\infty} dt~|j_1(Z, t)|}~.
\end{eqnarray}
As a result, one obtains
\begin{eqnarray} \label{eq: mean arrival}
\tau(Z) &=& \int_0^{\infty} dt~t ~ \Pi(Z, t)~,
\end{eqnarray}
for the mean arrival time at the detector location $z = Z$. It may
be noted that the above expression for the mean arrival time follows
uniquely \cite{unique} in a causal interpretation of quantum
mechanics provided by the Bohmian model \cite{bohm}. Note also that
eqs.(\ref{eq: arrival distribution}) and (\ref{eq: mean arrival})
represent the single-particle arrival time distribution and the mean
arrival time, respectively, when the effects of particle statistics
of the considered two-body system have been taken into account in
the computation of the single-particle current density through the
process of symmetrization or asymmetrization of the wavefunction. In
the following analysis we will consider the initial single particle
wavefunctions to be  Gaussians given by
\begin{eqnarray} \label{eq: initial_1p_Gauss_packet}
\psi_i(z, 0) = (2\pi \sigma_{0i}^2)^{-1/4} \exp{ \left[ {i k_i
z-\frac{(z-z_{ci})^2}{4\sigma_{0i}^2}} \right] }~, ~~~~~~~i=a, b
\end{eqnarray}
and then compute the arrival time distribution of the cases of first
free evolution, and then free fall under gravity. In eq.(\ref{eq:
initial_1p_Gauss_packet}), $\sigma_{0i}$, $z_{ci}$ and $k_i$
represent respectively the rms width, center and kick momentum of
the i-th packet.

%======================================================================

%\subsection{Two Gaussian wavepackets in a uniform gravitational filed}

\section{Effect of statistics on freely propagating wavepackets}

In this section we study the effect of symmetrization and asymmetrization of
a two-body freely propagating wavefunction on the arrival time distribution
computed through the probability current approach. The initial  one-particle
 wavefunctions are taken to be  Gaussian
packets given by eq.(\ref{eq: initial_1p_Gauss_packet}). The time
evolved wave functions are obtained to be
\begin{eqnarray} \label{eq: Gauss_packet0}
\psi_i(z, t) &=& \frac{1}{(2\pi s_{ti}^2)^{1/4}} \exp{ \left \{
\frac{im}{2 \hbar t} \left[ \left( z^2 +i\frac{\hbar t}{2 m
\sigma_{0i}^2 } z_{ci}^2 \right) -\frac{\sigma_{0i}}{s_{ti}} \left(
z - z_{ci} - \frac{\hbar t}{m}k_i  + \frac{s_{ti}}{ \sigma_{0i} }
z_{ci} \right)^2 \right] \right \} }~,
\end{eqnarray}
where $s_{ti} = \sigma_{0i}(1+i\hbar t/2m\sigma_{0i}^2)$ denotes the
time-evolved spread of the wavepackets. The overlap integral is
given by
\begin{eqnarray} \label{eq: innner-product-psia-psib_freeevolv}
\langle \psi_a(t) | \psi_b(t) \rangle = \sqrt{ \frac{2 \sigma_{0a}
\sigma_{0b} }{ \sigma_{0a}^2 + \sigma_{0b}^2 } }
~\exp{\left[-\frac{4(k_a-k_b)^2\sigma_{0a}^2 \sigma_{0b}^2 +
(z_{ca}-z_{cb})^2 + 4i(k_a-k_b)(\sigma_{0b}^2 z_{ca} + \sigma_{0a}^2
z_{cb})}{4(\sigma_{0a}^2 + \sigma_{0b}^2)} \right]} ~,
\end{eqnarray}
and thus, the normalization constants read
\begin{eqnarray} \label{eq: normalization-constants}
N_{\pm} &=& \frac{1}{\sqrt{2}}\left\{ 1 \pm \frac{2\sigma_{0a}
\sigma_{0b}}{\sigma_{0a}^2 + \sigma_{0b}^2}
\exp{\left[-\frac{4(k_a-k_b)^2\sigma_{0a}^2 \sigma_{0b}^2 +
(z_{ca}-z_{cb})^2}{2(\sigma_{0a}^2 + \sigma_{0b}^2)} \right]}
\right\}^{-1/2}~.
\end{eqnarray}
One sees that the magnitude of the overlap integral decreases
exponentially with the separation of the initial wavepackets. As the
separation grows, the  quantum and classical results approach:
\begin{eqnarray}
\rho_1(z, t) &\simeq & \frac{1}{2} \left(~|\psi_a(z, t)|^2 + |\psi_b(z, t)|^2 \right) = \rho_{\text{MB}}(z, t)~, \nonumber \\
j_1(z, t) &\simeq& \frac{1}{2} \frac{\hbar}{m} \Im \left \{
\psi^*_a \frac{\partial \psi_a}{\partial z} + \psi^*_b
\frac{\partial \psi_b}{\partial z} \right \} = j_{\text{MB}}(z, t)~.
\end{eqnarray}

The position expectation value with respect to $\rho_1(z, t)$ reads
\begin{eqnarray} \label{eq: z_exp rho_sp}
\langle z \rangle &=& \int z\rho_1(z, t)~dz
\nonumber \\
&=& |N_{\pm}|^2
\left \{
z_{\mathrm{cl},a} + z_{\mathrm{cl},b}
\pm |\langle \psi_a(t) | \psi_b(t) \rangle|^2
\left(
\frac{ \hbar t (k_a \sigma_{0a}^2 + k_b \sigma_{0b}^2) -
 m (z_{ca} \sigma_{0a}^2 + z_{cb} \sigma_{0b}^2) }{ m ( \sigma_{0a}^2 + \sigma_{0b}^2) }
\right)
\right \}
~,
\end{eqnarray}
where $z_{\mathrm{cl},a}$ denotes the classical path of the center of
the wavepacket $\psi_a$, i.e., $z_{\mathrm{cl},a} = z_{ca} + \hbar
k_a t/m$. One sees that the expectation value of
the center of mass operator is just the position expectation value
with respect to $\rho_1(z, t)$, i.e.,
\begin{eqnarray}
\langle z_{\text{cm}} \rangle(t) &=& \left\langle \frac{z_1+z_2}{2} \right \rangle(t)
= \int \int dz_1 dz_2 \frac{z_1+z_2}{2} |\Psi(z_1, z_2, t)|^2
\nonumber
\\
&=& \frac{1}{2} \left \{ \int dz_1 z_1\int dz_2 |\Psi(z_1, z_2,
t)|^2 + \int dz_2 z_2\int dz_1 |\Psi(z_1, z_2, t)|^2 \right \} =
\int dz~z~\rho_1(z, t) ~.
\end{eqnarray}
%
%
%********************************************************
% Figure
\begin{figure}
\centering
\includegraphics[width=12cm,angle=-90]{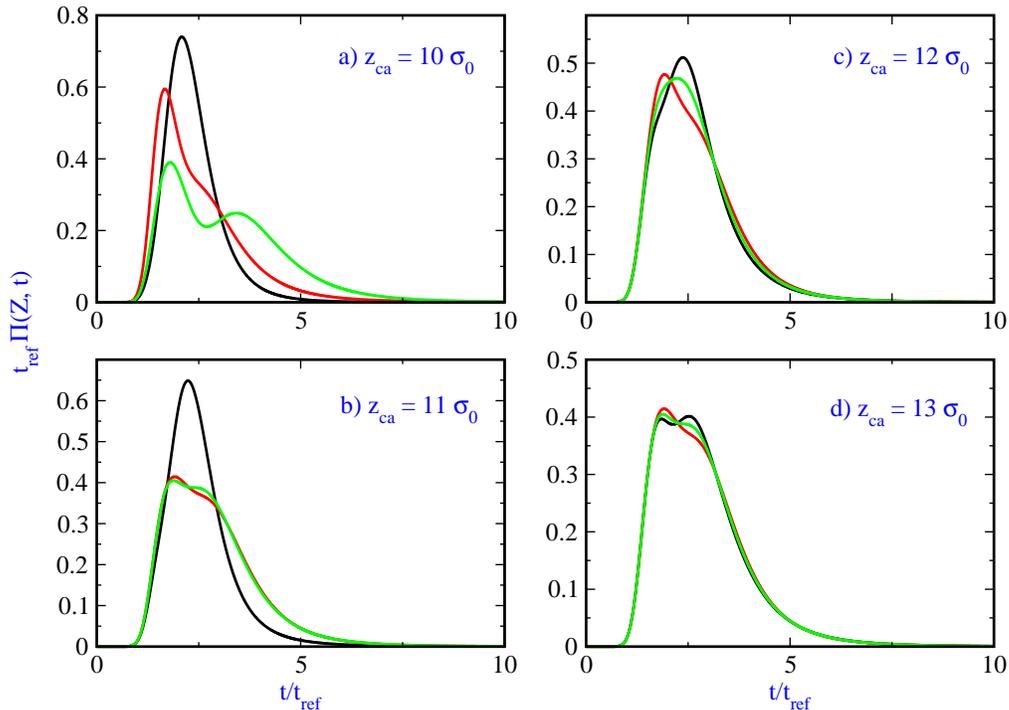}
\caption{(Color online) Arrival time distribution at
the detector location $Z=0$ for different initial separation of the
wavepackets for the free evolution case. Black curves stand for BE
statistics, the red ones are for the FD statistics and the green
ones show MB statistics. Here, $k_a = k_b = -2 / \sigma_0$, $z_{cb}
= 8 \sigma_0$, $\sigma_0 = 10~\mu$m, $m = m_n$ and
$t_{\text{ref}} = 2 m_{\text{n}} \sigma_0^2/\hbar = 3.165 $ ms.}
\label{fig: ardis_separation_freeevolve}
\end{figure}
%********************************************************

  We calculate the arrival time distribution at the detector location
$Z=0$ by substituting eq.(\ref{eq: Gauss_packet0}) in the expression
for the probability current given by eq.(\ref{eq: prob_qm_twobody}),
and then substituting the obtained expression in eq.(\ref{eq:
arrival distribution}). The values of the parameters used for our
numerical computations are given by: $\sigma_{0a} = \sigma_{0b} =
\sigma_{0} =  10~ \mu$m, $z_{ca} = 10\sigma_0$, $z_{cb} =
8\sigma_0$, $k_a = k_b = -2/\sigma_{0}$, $ m = m_{\text{n}} = 1.67
\times 10^{-27}~$kg, and $t_{\text{ref}} = \frac{2 m_{\text{n}}
\sigma_0^2}{\hbar} = 3.165 $ ms. In figure \ref{fig:
ardis_separation_freeevolve} the arrival time distribution has been
plotted for different values of initial separation of the
wavepackets. It is observed that the inclusion of particle
statistics through symmetrization and antisymmetrization of a
two-body wavefunction has a perceptible impact on the arrival time
distribution.
Such an effect of particle statistics on arrival times of freely
propagating particles has been studied earlier by using the
concept of the crossing state and the second quantization formalism
\cite{BaEgMu-PRA-2002}.
In the present work we further compute the effect of initial separation of the overlapping wavepackets, as is
displayed in the Figure \ref{fig:
ardis_separation_freeevolve}. We fix the center of $\psi_b$ and change
the center of $\psi_a$ from one plot to the other.
Using eq. (\ref{eq: innner-product-psia-psib_freeevolv}) and
eq. (\ref{eq: mean arrival}), for the parameters of
Fig.~\ref{fig: ardis_separation_freeevolve} one obtains
\begin{center}
 \begin{tabular}{ || c | c | c | c | c || }
   \hline
   $ \quad z_{ca} / \sigma_0 \quad $ & $ \quad \langle \psi_a | \psi_b \rangle \quad $ & $  \quad  \tau_{BE} / t_{ref} \quad $ & $  \quad \tau_{FD}/ t_{ref} \quad $ & $  \quad  \tau_{MB} / t_{ref} \quad $ \\ \hline
   10 & 0.6065 & 2.371 & 2.546 & 2.427 \\ \hline
   11 & 0.3247 & 2.518 & 2.614 & 2.561 \\ \hline
   12 & 0.1353 & 2.682 & 2.707 & 2.694\\ \hline
   13 & 0.04394 & 2.826 & 2.829 & 2.822 \\ \hline
 \end{tabular}
\end{center}
One sees that as the separation of the initial wavepackets grows, the quantum
and classical results approach as a result of decrement of the
overlap integral.

%===================================================
\section{Violation of the weak equivalence principle of quantum mechanics}

\subsection{Effect of statistics on freely falling wavepackets}

The arrival time of freely falling wavepackets using the probability current
approach has been studied earlier
\cite{AlMaHoPa-CQG-2006, ChHoMaMoMoSi-CQG-2012}, where it was noted that the
 the arrival time distribution acquires a mass-dependence due to wavepacket
spread. In the previous section we have seen that the effect of particle
statistics impacts the arrival time distribution even for the case of free
evolution. Now, let us study the effect of symmetrization and asymmetrization
of the wavefunction of a system of two particles falling freely under gravity.
Here again we choose the initial single-particle wavefunctions as Gaussians
given by eq.(\ref{eq: initial_1p_Gauss_packet}), and set the initial velocity
of the particles to be zero, i.e., the particles are dropped from rest with
 $k_a = k_b = 0$, and accelerate downwards under gravity with $g=10~$m/s$^2$.
All numerical calculations have been done based on the approximate
value $g=10$ m$/$s$^2$, since  using the exact value of $g$ is not
important for the arguments and reasoning.
The time-evolved single-particle wavefunctions in the uniform gravitational
field $ V(z) = m g z$ are given by,
\begin{eqnarray} \label{eq: Gauss_packet}
\psi_i(z, t) &=& \frac{1}{(2\pi s_{ti}^2)^{1/4}} \exp{ \left \{ \frac{im}{2 \hbar t} \left[ \left( z^2 - g t^2 z- \frac{g^2 t^4}{12}
+i\frac{\hbar t}{2 m \sigma_{0i}^2 } z_{ci}^2
\right)
-\frac{\sigma_{0i}}{s_{ti}} \left( z - z_{ci} + \frac{g t^2}{2} + \frac{s_{ti}}{ \sigma_{0i} } z_{ci} \right)^2
\right] \right \} }~,
\end{eqnarray}
where $s_{ti} = \sigma_{0i}(1+i\hbar t/2m\sigma_{0i}^2)$. The
overlap integral is given by,
\begin{eqnarray} \label{eq: innner-product-psia-psib}
\langle \psi_a(t) | \psi_b(t) \rangle = \sqrt{ \frac{2 \sigma_{0a}
\sigma_{0b} }{ \sigma_{0a}^2 + \sigma_{0b}^2 } }
~\exp{\left[-\frac{
(z_{ca}-z_{cb})^2}{4(\sigma_{0a}^2 + \sigma_{0b}^2)} \right]} ~,
\end{eqnarray}
and hence, the normalization constants become
\begin{eqnarray} \label{eq: normalization-constants}
N_{\pm} &=& \frac{1}{\sqrt{2}}\left\{ 1 \pm \frac{2\sigma_{0a}
\sigma_{0b}}{\sigma_{0a}^2 + \sigma_{0b}^2}
\exp{\left[-\frac{
(z_{ca}-z_{cb})^2}{2(\sigma_{0a}^2 + \sigma_{0b}^2)} \right]}
\right\}^{-1/2}~.
\end{eqnarray}
One sees that both overlap integral and normalization constants are
independent of $g$. Here again the  magnitude of the overlap
integral decreases exponentially with the separation of the initial
wavepackets, and as in the case of freely evolving wavepackets, the
quantum results of the probability density and current approach the
classical expressions as the  separation grows. The position
expectation value is given by
\begin{eqnarray} \label{eq: z_exp rho_sp}
\langle z \rangle_{\pm} &=& \int z\rho_1(z, t)~dz
\nonumber \\
&=& |N_{\pm}|^2 \left \{ z_{\mathrm{cl},a} + z_{\mathrm{cl},b} \pm
|\langle \psi_a(t) | \psi_b(t) \rangle|^2 \left( \frac{- g t^2}{2} -
\frac{z_{ca} \sigma_{0a}^2 + z_{cb} \sigma_{0b}^2 }{ \sigma_{0a}^2 +
\sigma_{0b}^2 } \right) \right \} ~,
\end{eqnarray}
where $z_{\mathrm{cl},a}$ shows the classical path of the center of
the wavepacket $\psi_a$, i.e, $z_{\mathrm{cl},a} = z_{ca}  - g t^2/2$. The
position expectation value
with respect to $\rho_1(z, t)$ is the expectation value of
the center of mass operator.
The quantity $\langle z \rangle _{\pm} $ is independent of
mass.

For the free fall case and for $ \sigma_{0a} = \sigma_{0b} = \sigma_0 $ one obtains
\begin{eqnarray}
\Delta z_{\pm}(t) &=& \sqrt{ \int dz~ z^2 \rho_1(z,t) - \left(  \int dz~ z~\rho_1(z,t)  \right)^2 }~, \nonumber \\
&=& \sigma_0 \sqrt{ \left( 1 + \frac{ \hbar^2 t^2 }{ 4 m^2 \sigma_0^4 }  \right)
\left \{ 1 \pm \frac{ (z_{ca}-z_{cb})^2 }{ 4 \sigma_0^2 \left( \mp 1 - \exp \left[ \frac{(z_{ca}-z_{cb})^2}{4 \sigma_0^2} \right] \right)} \right \} + \frac{ (z_{ca}-z_{cb})^2 }{4} }~,
\end{eqnarray}
for the width of the single-particle distribution. One clearly sees
this quantity is independent of the gravity strength $g$ and its
growth with time is slow for large masses.
%
%********************************************************
% Figure
\begin{figure}
\centering
\includegraphics[width=10cm,angle=-90]{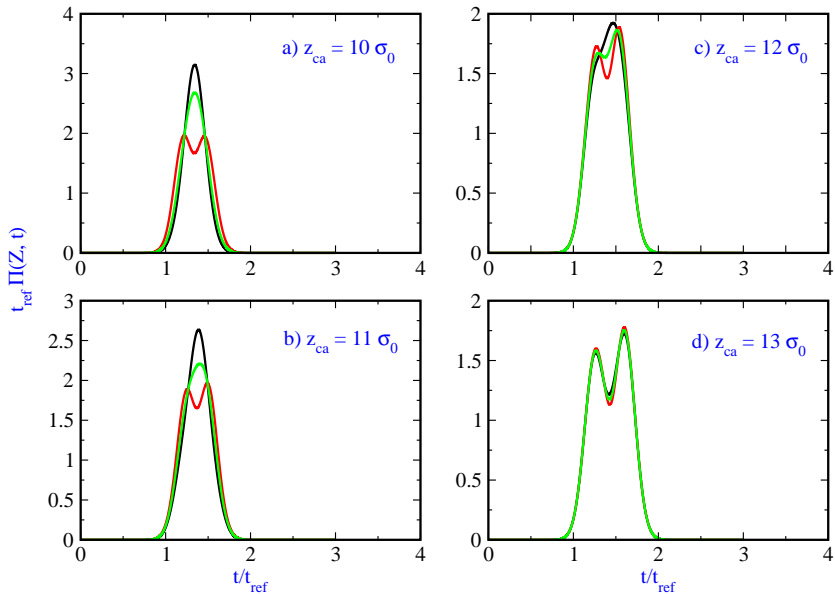}
\caption{(Color online) Arrival time distribution at
the detector location $Z=0$ for different initial separation of the
wavepackets for the free fall case. Black curves show BE statistics,
the red ones are for the FD statistics and the green ones show MB
statistics. Here,  $z_{cb} = 8 \sigma_0$, $\sigma_0 = 10~\mu$m, $g =
10~$m/s$^2$, $m = m_n$ and $t_{\text{ref}} = 2
m_{\text{n}} \sigma_0^2/\hbar = 3.165 $ ms.} \label{fig:
ardis_separation_freefall}
\end{figure}

%********************************************************
%
 We compute the arrival time distribution at the detector location
$Z=0$ by substituting the expression for the time-evolved wavefunction
eq.(\ref{eq: Gauss_packet}) in the expression for the probability
current given by eq.(\ref{eq: prob_qm_twobody}), and then
using eq.(\ref{eq: arrival distribution}).
Using the three quantities $\hbar$, $g$ and $\sigma_0$, one
can construct a quantity $ m_0 = \hbar/\sqrt{g \sigma_0^3} $ with
the dimension of mass, which has the value $ m_0 = 1.055 \times
10^{-27} $kg $\approx m_n$, the neutron mass. We use $m_n$ to make
the mass $m$ dimensionless.
The values of the parameters used for our numerical calculations are given
by:
$\sigma_{0a} = \sigma_{0b} = \sigma_{0} =  10~ \mu$m, $z_{ca} = 10\sigma_0$,
$z_{cb} = 8\sigma_0$,  $ m = m_{\text{n}} = 1.67 \times
10^{-27}~$kg, and $t_{\text{ref}} = \frac{2 m_{\text{n}}
\sigma_0^2}{\hbar} = 3.165 $ms.
In figure \ref{fig: ardis_separation_freefall} the arrival time
distribution has been plotted for different values of initial
separation of the wavepackets. In this case too the particle
statistics has an impact on the arrival time distribution.
Using eq. (\ref{eq: innner-product-psia-psib}) and
eq. (\ref{eq: mean arrival}), for the parameters of
Fig.~\ref{fig: ardis_separation_freefall} one obtains
\begin{center}
 \begin{tabular}{ || c | c | c | c | c || }
   \hline
   $ \quad z_{ca} / \sigma_0 \quad $ & $ \quad \langle \psi_a | \psi_b \rangle \quad $ & $  \quad  \tau_{BE} / t_{ref} \quad $ & $  \quad \tau_{FD}/ t_{ref} \quad $ & $  \quad  \tau_{MB} / t_{ref} \quad $ \\ \hline
   10 & 0.6065 & 1.339 & 1.341 & 1.340 \\ \hline
   11 & 0.3247 & 1.374 & 1.375 & 1.374 \\ \hline
   12 & 0.1353 & 1.407 & 1.407 & 1.407 \\ \hline
   13 & 0.04394 & 1.439 & 1.439 & 1.439 \\ \hline
 \end{tabular}
\end{center}
Such an effect of symmetrization and asymmetrization  of a two-body
wavefunction on arrival times of freely falling wavepackets may be
regarded as nonlocal (in the sense that the single-particle arrival
time distribution depends on the spatially separated second
particle, as well), and thus contrary to the tenet of the local weak
equivalence principle of classical gravity, which forms the
inspiration of the statement of WEQ.

%

%********************************************************
% Figure
\begin{figure}
\centering
\includegraphics[width=10cm,angle=-90]{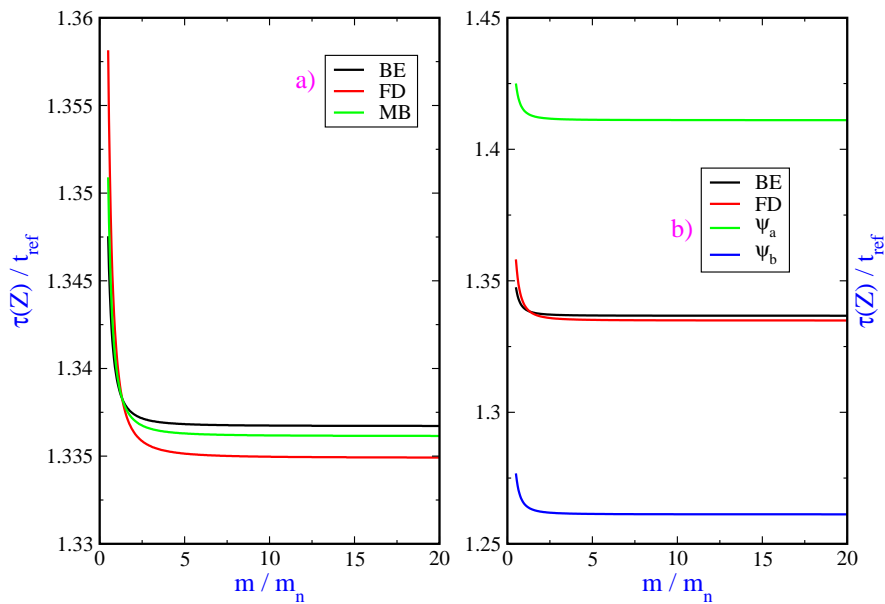}
\caption{(Color online) Mean arrival time at the
detector location $Z=0$ versus mass for the free fall case. Here,
$k_a = k_b = 0$, $z_{ca} = 10 \sigma_0$, $z_{cb} = 8 \sigma_0$,
$\sigma_0 = 10 ~\mu$m and $t_{\text{ref}} = 2
m_{\text{n}} \sigma_0^2/\hbar = 3.165 $ ms. In b) we have included
mean arrival time of single-particle systems which are described by
wavepackets $\psi_a$ and $\psi_b$, i.e., $\tau_a$ and $\tau_b$.}
\label{fig: freefall_tau_mass}
\end{figure}
%********************************************************

Now, in order to show the explicit mass dependence of arrival time
for particles in free fall we plot the mean arrival time versus mass
in the figure \ref{fig: freefall_tau_mass}. Here again, the mean
arrival time is computed at the detector location $Z=0$ using
eq.(\ref{eq: arrival distribution}) and eq.(\ref{eq: mean arrival})
after  substituting the expression for the time-evolved wavefunction
eq.(\ref{eq: Gauss_packet}) in the expression for the probability
current given by eq.(\ref{eq: prob_qm_twobody}). The values of the
parameters used are as before. Curves with label $\psi_a$ ($\psi_b$)
show mean arrival time for wavepackets $\psi_a$ ($\psi_b$)
respectively. The MB mean arrival time is given by $
\tau_{\text{MB}} = \frac{ \tau_a + \tau_b }{2}$. One sees that for
all types of statistics the mean arrival time decreases with mass at
first and then becomes constant for large mass. The effect of
symmetrization and asymmetrization of the wavefunction has a
perceptible effect on the magnitude of the violation of WEQ.
However, the violation of WEQ observed  explicitly for low masses
tends to disappear smoothly in the limit of large mass. Hence, even
with the inclusion of particle statistics,  the classical limit
emerges smoothly through this approach.

\begin{figure}
\centering
\includegraphics[width=10cm,angle=0]{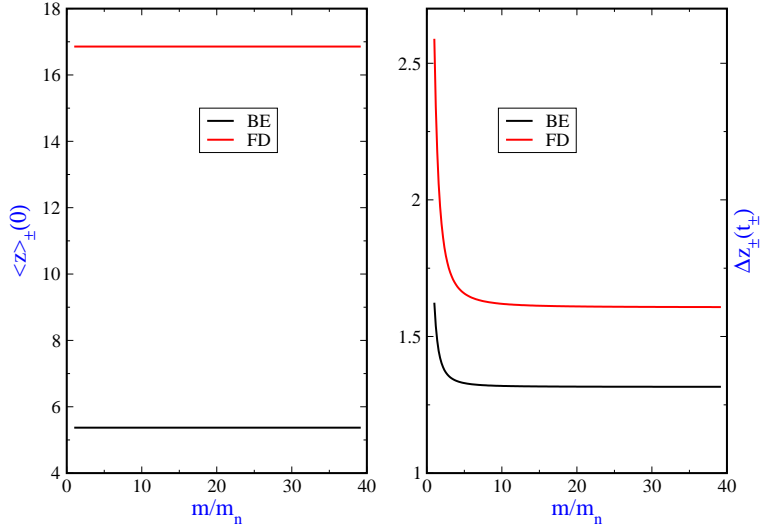}
\caption{(Color online)  $\langle z \rangle_{\pm}(0)$ and $\Delta
z_{\pm}(t_{\pm})$ for the free fall case. Black curve shows BE
statistics and the red one is for the FD statistics. Here,  $z_{ca}
= 10 \sigma_0$ $z_{cb} = 8 \sigma_0$, $\sigma_0 = 10~\mu$m and $g =
10~$m/s$^2$.} \label{fig: zDelz_mass}
\end{figure}

One sees each of the three statistics gives a
different mean arrival time in the large mass limit. The difference
between the mean arrival times of bosons and fermions is related to
the difference between the two quantities (i) position expectation
value, $\langle z \rangle(t)$, with respect to the one-body
probability density $\rho_1(z, t)$ and (ii) spread in position,
$\Delta z(t) = \sqrt{ \langle z^2 \rangle - \langle z \rangle^2 }$.
At time $ t= t_{\pm} $,
\begin{eqnarray}
t_{\pm} = \sqrt{ \frac{ z_{ca} + z_{cb} }{g} ~ \frac{ 2 \mp \exp
\left[- \frac{(z_{ca}-z_{cb})^2}{4 \sigma_0^2} \right] }{ 2 \pm \exp
\left[- \frac{(z_{ca}-z_{cb})^2}{4 \sigma_0^2} \right] } }~.
\end{eqnarray}
the center of the one-body probability density $\rho_1(z, 0)$ arrives at
the ground, the detector location.
We have plotted $\langle z \rangle_{\pm}(0)$ and $\Delta
z_{\pm}(t\pm)$ as a function of mass in Fig.~\ref{fig: zDelz_mass}.
As this figure shows  $\Delta z_{+}(t_+)$ and $\Delta z_{-}(t_-)$
corresponding to bosonic and fermionic statistics, respectively, do
not approach each other even in the large mass limit.
%
%********************************************************
% Figure

%********************************************************
%

\subsection{Spin-dependent arrival time distribution}

In this subsection we discuss the effect of explicit inclusion of
spin in the probability current density on the arrival time
distribution of particles in free fall. It has been noted earlier
\cite{fink} that in nonrelativistic quantum mechanics the form of
the probability current density is not unique. It was shown
\cite{Ho-PRA-1999} that the Dirac equation implies a unique
expression for the probability current for spin-$1/2$ particles in
the nonrelativistic limit. Uniqueness of the probability current is
a generic consequence of any relativistic quantum dynamics. Effects
of spin on the arrival time distribution for freely evolving
wavepackets have been computed for fermions \cite{spin1} as well as
for bosons \cite{struyve}. Here we obtain the spin-dependent arrival
time distribution for a symmetric two-dimensional Gaussian
wavepacket in order to compare the effect of explicit presence of
spin in the probability current density with the effect of particle
statistics for wavepackets falling freely under gravity. Taking into
account the spin contribution to the probability current density for
particles with spin-$1/2$, one obtains \cite{Ho-PRA-1999}
\begin{eqnarray}
{\bf{j}}({\bf{x}}, t; {\bf{\hat{s}}}) &=&
{\bf{j}}_{\text{Sch}}({\bf{x}}, t) + \frac{1}{m} \nabla
[\psi^*({\bf{x}}, t) \psi({\bf{x}}, t)] \times {\bf{s}}
\\
&=& \frac{\hbar}{m} \{ \Im[ \psi^*({\bf{x}}, t) \nabla
\psi({\bf{x}}, t) ] + \Re[ \psi^*({\bf{x}}, t) \nabla \psi({\bf{x}},
t) ] \times {\bf{\hat{s}}} \} ~,
\end{eqnarray}
for a spin eigenstate $ \chi $ in the absence of a magnetic field.
Here, $ {\bf{j}}_{\text{Sch}}({\bf{x}}, t) $ is the usual
Schr\"odinger current, whereas the second term denotes the
spin-dependent contribution with $ {\bf{s}} =
\frac{\hbar}{2}{\bf{\hat{s}}} = \frac{\hbar}{2}
\chi^{\dagger}{\bf{\hat{\sigma}}} \chi $.
For a two-dimensional system with a wavefunction which is in the
factorized form $\psi({\bf{x}}, t) = \psi_x(x, t) \psi_z(z, t)$, the
probability current density takes the form
\begin{eqnarray}
j_x(x, z, t; {\bf{\hat{s}}}) &=& \frac{\hbar}{m} \{ |\psi_z(z, t)|^2
\Im[ \psi_x^*(x, t) \partial_x \psi_x(x, t) ] - |\psi_x(x, t)|^2
\Re[ \psi_z^*(z, t) \partial_z \psi_z(z, t) ] \} ~,
\\
j_z(x, z, t; {\bf{\hat{s}}}) &=& \frac{\hbar}{m} \{ |\psi_x(x, t)|^2
\Im[ \psi_z^*(z, t) \partial_z \psi_z(z, t) ] + |\psi_z(z, t)|^2
\Re[ \psi_x^*(x, t) \partial_x \psi_x(x, t) ] \} ~,
\end{eqnarray}
for $ {\bf{\hat{s}}} = (0, 1, 0)$. In the uniform gravitational
field $ V(z) = m g z $, by taking the initial wavefunction as a
two-dimensional Gaussian wavepacket,
\begin{eqnarray} \label{2Dsymmetric_Gaussian}
\psi_0(x, z) &=& \frac{1}{\sqrt{2\pi} \sigma_0} \exp \left \{
-\frac{x^2}{4 \sigma_0^2} - \frac{(z-z_c)^2}{4 \sigma_0^2} + i k_0 z
\right \}~,
\end{eqnarray}
the modulus of the Schr\"{o}dinger current takes the form
\begin{eqnarray} \label{eq: j_sch_symmetric}
|{\bf{j}}_{\text{Sch}}(x, z, t)| &=& \frac{ \sqrt{ f_4(t) m^4 +
f_3(t) m^3 + f_2(z, t) m^2 + f_1(z, t) m + f_0(x, z, t) } } {16 \pi
m^2  \sigma_0^2 \sigma_t^4}
\nonumber \\
& \times & \exp \left \{-\frac{  x^2 + [ z - z_{\text{cl}}(t) ]^2
}{2 \sigma_t^2} \right \} ~,
\end{eqnarray}
while the modulus of the total spin-dependent probability current
density is given by
\begin{eqnarray} \label{eq: j_total_symmetric}
|{\bf{j}}(x, z, t; {\bf{\hat{s}}})| &=& \frac{ \sqrt{ h_2(t) m^2 +
h_1(x, t) m + h_0(x, z, t) } } {8 \pi m \sigma_0 \sigma_t^3} \exp
\left \{-\frac{  x^2 + [ z - z_{\text{cl}}(t) ]^2 }{2 \sigma_t^2}
\right \} ~,
\end{eqnarray}
where,
\begin{eqnarray*}
f_0(x, z, t) &=& \hbar^4 t^2 \{ g^2 t^4 + 4 [x^2 + (z - z_c)^2] + 4 g t^2 (-z+z_c) \}~, \\
f_1(z, t) &=&  -16 \hbar^3 k_0 \sigma_0^4 t (g t^2 - 2 z + 2 z_c)~, \\
f_2(z, t) &=& 16 \hbar^2 \sigma_0^4 [4 k_0^2 \sigma_0^4 + g t^2 (g t^2 - 2 z + 2 z_c)]~, \\
f_3(t) &=& - 128 g \hbar k_0 \sigma_0^8 t~, \\
f_4(t) &=& 64 g^2 \sigma_0^8 t^2~,
\end{eqnarray*}
and
\begin{eqnarray*}
h_0(x, z, t) &=& \hbar^2 \{ 16 k_0^2 \sigma_0^4 + g^2 t^4 - 16 k_0 \sigma_0^2 x + 4[ x^2 + (z - z_c)^2 ] + 4 g t^2 (- z + z_c) \}~,\\
h_1(x, t) &=& 16 g \hbar \sigma_0^2 t (-2 k_0 \sigma_0^2 +x)~, \\
h_2(t) &=& 16 g^2 \sigma_0^4 t^2~.
\end{eqnarray*}
In eqs. (\ref{eq: j_sch_symmetric}) and (\ref{eq:
j_total_symmetric}), $\sigma_t = \sigma_0 \sqrt{ 1 + \frac{\hbar^2
t^2}{4m^2 \sigma_0^4} }$ and $ z_{\text{cl}}(t) = -\frac{1}{2}gt^2 +
u_z t + z_c $ with $ u_z = \hbar k_0/m $.

In order to obtain the arrival time distribution at the {\it plane}
$ z=Z $ we integrate over the space coordinate $x$. Thus, the {\it
normalized} arrival time distribution is given by
\begin{eqnarray} \label{eq: sch arr_dis}
\Pi_{\text{Sch}}(Z, t) &=& \frac{ \int_{-\infty}^{\infty}
dx~|{\bf{j}}_{\text{Sch}}(x, Z, t)| } { \int_0^{\infty}
dt\int_{-\infty}^{\infty} dx~|{\bf{j}}_{\text{Sch}}(x, Z, t)| }
\end{eqnarray}
in the absence of spin contribution, while it reads
\begin{eqnarray} \label{eq: spin_dep arr_dis}
\Pi(Z, t; {\bf{\hat{s}}}) &=& \frac{ \int_{-\infty}^{\infty}
dx~|{\bf{j}}(x, Z, t; {\bf{\hat{s}}})| } { \int_0^{\infty}
dt\int_{-\infty}^{\infty} dx~|{\bf{j}}(x, Z, t; {\bf{\hat{s}}})| }~,
\end{eqnarray}
considering the spin effect. We have integrated over $ x $ from
$-\infty$ to $\infty$ to compare the results with the corresponding
one-dimensional system. The corresponding mean arrival times are
respectively given by
\begin{eqnarray} \label{eq: mean_arriv_times}
\tau_{\text{Sch}}(Z) &=& \int_0^{\infty} dt~t~ \Pi_{\text{Sch}}(Z, t)~, \\
\tau(Z; {\bf{\hat{s}}}) &=& \int_0^{\infty} dt~t ~\Pi(Z, t; {\bf{\hat{s}}})~.
\end{eqnarray}
%
%********************************************************
% Figure
\begin{figure}
\centering
\includegraphics[width=10cm,angle=-90]{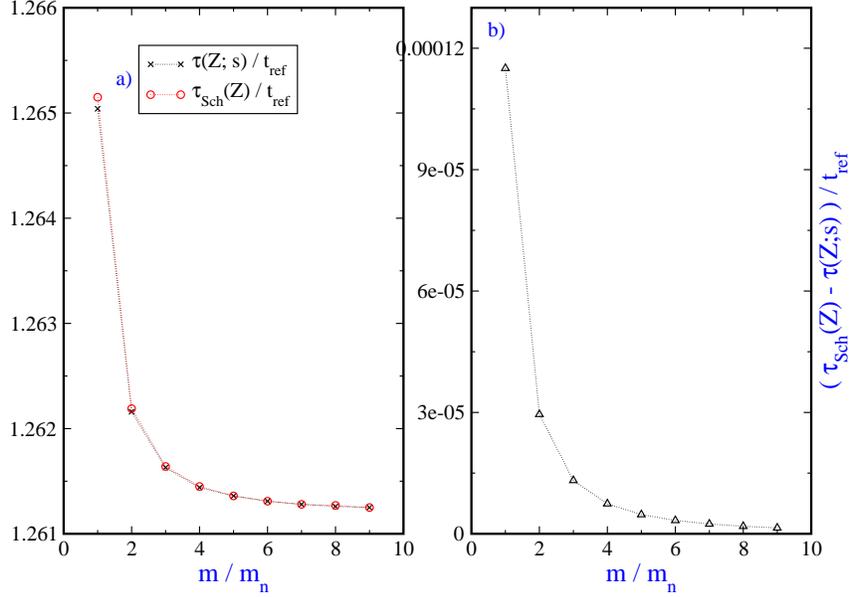}
\caption{(Color online) a) Spin-dependent (black
points) and Spin-independent (red points) mean arrival time at the
detector plane $Z=0$ and b) their difference, versus mass for the
free fall of the 2D symmetric Gaussian packet
(\ref{2Dsymmetric_Gaussian}). The parameters of the falling
wavepacket have been chosen as $z_c = 8 \sigma_0$ and
$\sigma_0=10~\mu$m,  $ m_{\text{n}} = 1.67 \times 10^{-27}~$kg and
$t_{\text{ref}} = 2 m_n \sigma_0^2/\hbar = 3.161~$ms. } \label{fig:
freefall_meanmass}
\end{figure}
%********************************************************

In the Figure \ref{fig: freefall_meanmass}, the mean arrival times
$\tau_{\text{Sch}}$ and $\tau({\bf{\hat{s}}})$ at detector location
$ Z=0 $ have been plotted versus mass  for free fall from rest (
$k_0 = 0$). The parameters of the initial wavepacket have been
chosen as $z_c = 8 \sigma_0$ and $\sigma_0=10~\mu$m,  $ m_{\text{n}}
= 1.67 \times 10^{-27}~$kg and $t_{\text{ref}} = \frac{2 m_n
\sigma_0^2}{\hbar} = 3.161~$ms. One sees that the mean arrival time
is mass-dependent, thus signifying a violation of WEQ. Inclusion of
spin causes the mean arrival time to decrease with a small magnitude
compared to the case of no spin. Hence, in comparison with the
results  for the case considering the effect of particle statistics
without spin-dependence displayed in the figure \ref{fig:
freefall_tau_mass}, we find that the contribution of including spin
explicitly at the level of the probability current leads to  a much
smaller modification of the mean arrival time values. However, here
(Figure  \ref{fig: freefall_meanmass}) again we find that the mass
dependence of the mean arrival time vanishes for large mass,
signifying emergence of WEQ in the classical limit.

%======================================================

\section{Conclusions}

In this work we have investigated the effect of particle statistics on
the arrival time distribution of Gaussian wave packets. We have employed
the probability current approach \cite{leavens} for computing
the arrival time distribution
for a system of two particles. The probability current approach provides
an unambiguous definition \cite{leavens,unique} of the arrival time of quantum
particles, and
also leads to a proper classical limit \cite{classlim} of quantum dynamics.
The single-particle arrival time distributions
using the probability current approach have been studied earlier in the
literature for the cases of free evolution, leading to the interesting effect
of the spin-dependence \cite{Ho-PRA-1999,spin1,struyve} of arrival time. It
was also observed earlier that
the mean arrival time for particles freely falling under gravity acquires
mass-dependence, thus signifying violation of the weak equivalence principle
of quantum mechanics  \cite{AlMaHoPa-CQG-2006,ChHoMaMoMoSi-CQG-2012}.
Such  violation stems essentially from the spread of
propagating wavepackets that is mass-dependent.

In the present analysis we have shown for the first time that
symmetrization or asymmetrization of a two body wavefunction leads
to modification of the arrival time distribution even for freely
propagating wavepackets. The impact of particle statistics on the
arrival time distribution is observed for initially overlapping
Gaussian wavepackets. For the case of particles falling freely under
gravity, the mean arrival time acquires mass dependence signifying
violation of WEQ. The magnitude of violation is seen to be different
for MB, BE and FD statistics as a consequence of the difference in
the mass dependence of the mean arrival time for the case of
different statistics. We have also compared such a violation under
the influence of particle statistics with the violation of WEQ
obtained through the explicit introduction of particle spin in the
probability current. The modification of the mean arrival time in
the former approach is shown to be larger than in the latter for the
case of Gaussian wavepackets considered here. However, in both the
approaches, WEQ emerges smoothly in the limit of large mass.
It may be noted that each of the three statistics gives a different mean arrival time in the large mass limit. Such
 difference stems essentially from the difference in the position
expectation value and the spread of the wave packets for the
respective statistics, and may be viewed as a remnant effect of
particle statistics in the classical limit.

We conclude by observing that the effect of particle statistics in different
quantum phenomena is regarded to be of much importance motivating newer tests
beyond traditional arenas \cite{eng}. Our approach of studying the influence of
particle statistics on the arrival time distribution provides an independent
 manifestation of the fundamental quantum property of indistinguishability.
Moreover, the analysis of transit and flight times \cite{tof1,tof2}
is a key ingredient in experiments involving Bose condensates which
form a major avenue of revealing the effect of quantum statistics.
The results reported in the present paper using symmetric Gaussian
wavepackets furnish an example of the effect
of particle statistics on the arrival time distribution.
 Though the magnitude of the difference in observable
quantities resulting from different statistics, as computed in the
present work may not be high enough for direct experimental detection,
they provide an {\it in principle} signature of differential
violation of WEQ for different statistics.
More elaborate calculations are needed using different sets of parameters, and
various forms of wavepackets such as asymmetric \cite{spin1} or
non-Gaussian \cite{ChHoMaMoMoSi-CQG-2012} ones in order
 to raise the possibility of observational verification
of such  effects.

\vspace{5mm}
{\emph Acknowledgements:} ASM and DH acknowledge support from the
project SR/S2/LOP-08/2013 of DST, India. DH acknowledges support
from the Centre for Science, Kolkata.

%===============================================================

%********************************************************

%===========================
\end{document}